\documentclass[12pt]{article}
\begin{document}
\begin{center}
{\large{\bf Evolution of simple configurations of gravitating gas}}

\vspace*{10mm}

{\bf G.P. Pronko}\footnote{e-mail:pronko@ihep.ru}\\
{\it IHEP, Protvino, Moscow reg., Russia\\
}

\begin{abstract}
We considered the dynamics of gravitating gas - a continuous media
with peculiar properties. The exact solutions of its Euler equations
for simple initial conditions is obtained.
\end{abstract}
\end{center}

\section{Introduction}

Gravitating gas is a mechanical system which could be considered as a
continuous limit of the $N$-particles system described by Hamiltonian:
\begin{eqnarray}\label{1}
H=\sum_{i=1}^{N}\frac{\vec
p{_i}^2}{2m_i}-\sum_{i\not=j}^{N}\frac{G m_i m_j}{|\vec x_i-\vec
x_j|} ,
\end{eqnarray}
where $\vec x_i, \vec p_i$ are canonical coordinates of particles
(stars)
with masses $m_i$, $G$-gravitational constant.

When $N\rightarrow\infty$(with assumption $m_i=m$) the system turns into the
kind of continuous media with very peculiar properties due to attractive interaction
between constituents. In the usual gas the effective interaction between particles
(atoms or molecules) is  repulsive  and as a result being put in a
volume it spreads in space, filling after some time the whole volume
with
uniform density. On the opposite, as was pointed out in \cite{gravgas},
the gravitating gas, because of attractive interaction may form isolated steady
configuration with asymptotically vanishing density. In particularly,
the galaxy named Hoag's object may be an example of such
gravitational soliton \cite{gravgas}. Also, as we shall see below there is no
sound waves in gravitating gas and local perturbation of density in linear approximation
creates instability. That is why in this system in order to study evolution of perturbation
we need to take into account nonlinearity of the equations of motion.
In the present paper we shall consider the cases, where we can solve equations of motion exactly.

\section{Equations of motion}

In fluid (gas) mechanics there are two different pictures of
description.
The first, usually refereed as Eulerian, uses as the coordinates  the
space dependent fields of velocity and density. The second,
Lagrangian
description, uses the coordinates of the
particles $\vec x(\xi_i,t)$  labeled by the set of the parameters
$\xi_i$ (the numbers of particles in (\ref{1})), which could be
considered as the initial positions
$\vec\xi=\vec x(\xi_i,t=0)$  and time $t$. The useful physical
assumption is  that  the
functions $\vec x(\xi_i,t)$ define a diffeomorphism of  $R^3$
and the inverse functions $\vec \xi(x_i,t)$ should also exist.
\begin{eqnarray}\label{2}
x_j(\xi_i,t)\Big|_{\vec \xi=\vec\xi(x_i,t)}&=x_j,\nonumber\\
\xi_j(x_i,t)\Big|_{\vec x=\vec x(\xi_i,t)}&=\xi_j.
\end{eqnarray}
The density of the particles in space at time $t$ is
\begin{equation}\label{3}
\rho(\vec x,t)=\int d^3 \xi \rho_0(\xi_i)\delta(\vec x-\vec
x(\xi_i,t)),
\end{equation}
where $\rho_0(\xi)$ is the initial density at time $t=0$.
The velocity field  $\vec v$ as a function of coordinates $\vec x$
and $t$
is:
\begin{equation}\label{4}
\vec v(x_i,t)=\dot{\vec x}(\vec\xi(x_i,t),t),
\end{equation}
where $\vec\xi(x,t)$ is the inverse function (\ref{2}). The velocity
also
could be written in the following form:
\begin{equation}\label{5}
\vec v(x_i,t)=\frac{\int d^3 \xi \rho_0(\xi_i)\dot{\vec
x}(\xi_i,t)\delta(\vec
x-\vec x(\xi_i,t))}{\int d^3 \xi \rho_0(\xi_i)\delta(\vec x-\vec
x(\xi_i,t))},
\end{equation}
or
\begin{equation}\label{6}
\rho (x_i,t)\vec v(x_i,t)=\int d^3 \xi \rho_0(\xi_i)\dot{\vec
x}(\xi_i,t)\delta(\vec x-\vec x(\xi_i,t)).
\end{equation}
Let us calculate the time derivative of the density using its
definition (\ref{4}) :
\begin{eqnarray}\label{7}
&\dot \rho (x_i,t)=\displaystyle\int d^3 \xi
\rho_0(\xi_i)\frac{\partial}{\partial t}
\delta(\vec x-\vec x(\xi_i,t))\nonumber\\
&=\displaystyle\int d^3 \xi \rho_0(\xi_i)\left(-\dot{\vec
x}(\xi_i,t)\right)\frac{\partial}{\partial \vec x}
\delta(\vec x-\vec x(\xi_i,t))\nonumber\\
&=-\frac{\partial}{\partial \vec x}\int d^3 \xi
\rho_0(\xi_i)\dot{\vec x}(\xi_i,t)\delta(\vec x-\vec
x(\xi_i,t))\nonumber\\
&=-\displaystyle\frac{\partial}{\partial \vec x}\rho (x_i,t)
\vec v (x_i,t)
\end{eqnarray}
In such a way we verify the continuity equation of fluid dynamics:
\begin{equation}\label{8}
\dot \rho (x_i,t)+\vec\partial\Bigl(\rho (x_i,t)\vec v
(x_i,t)\Bigr)=0.
\end{equation}.

Using the coordinates $\vec x(\xi_i,t)$ as a configurational space
variables we can write the Lagrangian for the continuous
generalization of the system, described by (\ref{1})
\begin{equation}\label{9}
L=\int d^3 \xi \rho_0(\xi_i)\frac{m\dot{\vec x}^2
(\xi_i,t)}{2}+\frac{G m^2}{2} \int d^3 \xi d^3 \xi'
\frac{\rho_0(\xi_i)\rho_0(\xi'_i)}{|\vec x(\xi_i,t)-\vec
x(\xi'_i,t)|},
\end{equation}
 The equations of
motion, which follow from the Lagrangian (\ref{10}) have the form:
\begin{equation}\label{10}
m\ddot{\vec x} (\xi_i,t)+G m^2\int d^3 \xi'\rho_{0}(\xi'_i)
\frac{\vec x(\xi_i,t)-\vec x(\xi'_i,t)}{|\vec x(\xi_i,t)-\vec
x(\xi'_i,t)|^3}=0.
\end{equation}
Now we need to translate the equations (\ref{10})  into the language
of Euler variables. For this let us differentiate both sides of the
equation (\ref{6}) with respect to time:
\begin{eqnarray}\label{11}
&\displaystyle\frac{\partial}{\partial t}\rho (x_i,t)\vec
v(x_i,t)=\int
d^3 \xi \rho_0(\xi_i)\ddot{\vec x}(\xi_i,t)\delta(\vec x-\vec
x(\xi_i,t))\nonumber\\
&+\int d^3 \xi
\rho_0(\xi_i)\dot{\vec
x}(\xi_i,t)\displaystyle\frac{\partial}{\partial
t}\delta(\vec x-\vec x(\xi_i,t)).
\end{eqnarray}
Substituting $\ddot{\vec x}(\xi_i,t)$ from the equation (\ref{10})
and transforming the second term, as we did in derivation of the
equation (\ref{7})we obtain:
\begin{eqnarray}\label{12}
&\displaystyle  \frac{\partial}{\partial t}\rho (x_i,t)\vec
v(x_i,t)= -\displaystyle\frac{\partial}{\partial x_k}\Bigl(\rho
(x_i,t)
\vec v (x_i,t) v_k(x_i,t)\Bigr)+\nonumber\\
&\displaystyle\int d^3 \xi \rho_0(\xi_i)\Bigl[-\gamma\int d^3
\xi'\rho_{0}(\xi'_i)
\frac{\vec x(\xi_i,t)-\vec x(\xi'_i,t)}{|\vec x(\xi_i,t)-\vec
x(\xi'_i,t)|^3}\Bigr]\delta(\vec x-\vec
x(\xi_i,t))\nonumber\\,
\end{eqnarray}
where we introduced notation $\gamma=Gm$.
To transform the last integral we first perform integration over the
$\xi$:
\begin{eqnarray}\label{13}
&\displaystyle\int d^3 \xi \rho_0(\xi_i)\Bigl[-\gamma\int d^3
\xi'\rho_{0}(\xi'_i)
\frac{\vec x(\xi_i,t)-\vec x(\xi'_i,t)}{|\vec x(\xi_i,t)-\vec
x(\xi'_i,t)|^3}\Bigr]\delta(\vec x-\vec
x(\xi_i,t)) \nonumber \\
&=-\displaystyle \gamma \rho(x_i,t)\int d^3
\xi'\rho_{0}(\xi'_i)
\frac{\vec x-\vec x(\xi'_i,t)}{|\vec x-\vec
x(\xi'_i,t)|^3}.
\end{eqnarray}
Now let us insert in the integral over $\xi'$ the unity
\begin{equation}\label{14}
1=\int d^3 y \delta({\vec y- \vec x(\xi'_i,t)})
\end{equation}
and change the order of integration:
\begin{eqnarray}\label{15}
&\displaystyle\int d^3
\xi'\rho_{0}(\xi'_i)
\frac{\vec x-\vec x(\xi'_i,t)}{|\vec x-\vec
x(\xi'_i,t)|^3}\nonumber\\
&=\displaystyle\int d^3 y \int d^3
\xi'\rho_{0}(\xi'_i)
\frac{\vec x-\vec x(\xi'_i,t)}{|\vec x-\vec
x(\xi'_i,t)|^3}\delta({\vec y- \vec x(\xi'_i,t)})\nonumber\\
&\displaystyle=\int d^3 y
\rho(y_i,t)
\frac{\vec x-\vec y}{|\vec x-\vec
y|^3},
\end{eqnarray}
where in the last step we have used the definition of $\rho(y_i,t)$ given by (\ref{3}). Finally the equation of motion (\ref{11}) in terms of Euler variables takes
the following form:
\begin{eqnarray}\label{16}
&\displaystyle\frac{\partial}{\partial t}\rho (x_i,t)\vec
v(x_i,t)+\displaystyle\frac{\partial}{\partial x_k}\Bigl(\rho
(x_i,t)\vec v (x_i,t)
v_k(x_i,t)\Bigr)= \nonumber \\
&=\displaystyle \gamma\rho(x_i,t)\frac{\partial}{\partial
\vec x} \int d^3 y \frac{\rho (y_i,t)}{|\vec x-\vec y|}.
\end{eqnarray}
Using the continuity equation we can rewrite (\ref{16}) in the more
familiar  form of Euler equation:
\begin{eqnarray}\label{17}
\displaystyle\frac{\partial}{\partial t}\vec
v(x_i,t)+\displaystyle v_k(x_i,t)\frac{\partial}{\partial x_k}\vec v
(x_i,t)= \gamma\frac{\partial}{\partial
\vec x} \int d^3 y \frac{\rho (y_i,t)}{|\vec x-\vec y|}.
\end{eqnarray}

\section{Absence of sound waves in gravitating gas.}

As very well known (see for example \cite{Landau}), small
perturbation of density in gas or fluid (and/or velocity) creates
the sound waves. In the case of gravitating gas, due to attractive
interaction of constituents the situation is completely different.
Let us introduce a small perturbation of density
$\tilde{\rho}(x_i,t)$, so that
\begin{equation}\label{19}
\rho(x_i,t)=\rho_0+\tilde{\rho}(x_i,t),
\end{equation}
where $\rho_0(x_i)$ is background density. Neglecting nonlinear term in Euler equation we have
\begin{eqnarray}\label{20}
\dot{\vec v}(x_i,t)= \gamma\frac{\partial}{\partial
\vec x} \int d^3 y \frac{\rho (y_i,t)}{|\vec x-\vec y|}.
\end{eqnarray}
In linear approximation the continuity equation will have the
following form:
\begin{equation}\label{21}
\dot{\tilde{\rho}}(x_i,t)+\rho_0\frac{\partial v_k(x_i,t)}{\partial
x_k}=0.
\end{equation}
Taking time derivative of equation (\ref{21}) and substituting
$\dot{\vec v}(x_i,t)$ from  equation(\ref{20}) we obtain:
\begin{equation}\label{22}
\ddot{\tilde{\rho}}(x_i,t)+\rho_0\gamma \triangle\int d^3 y
\frac{\rho (y_i,t)}{|\vec x-\vec y|}=0.
\end{equation}
Action of Laplace operator on the potential gives the density times $-4\pi$ and finally we arrive at
\begin{equation}\label{23}
\ddot{\tilde{\rho}}(x_i,t)-4\pi\rho_0\gamma(\rho_0(x_i)+\tilde{\rho}(x_i,t))
=0.
\end{equation}
The solution of this equation has exponential time dependence which
makes an assumption of small perturbation invalid. Also this example
shows that the linear approximation does not work for gravitating
gas. The reason for such behavior of perturbation is the attraction
forces acting between constituents. In the case of usual gas with
repulsive interaction between constituents the r.h.s. of the
equation (\ref{20}) will stay e.g.
$-\frac{k^2}{\rho_0}\frac{\partial\rho(x_i,t)}{\partial \vec x}$ ,
where $k$ is the velocity of sound and therefore we will obtain the
instead of (22) the usual wave equation:
\begin{equation}\label{24}
\ddot{{\rho}}(x_i,t)+k^2\triangle\rho(x_i,t)) =0.
\end{equation}
The conclusion from this
observation is that if we want to study the behavior of gravitating gas, the
linear approximation is not sufficient and we need to solve Euler equations exactly.

\section{Exact evolution of spherically symmetric configuration of gravitating gas}

Here we will consider the spherically symmetric configuration of gravitating
gas which is described by $\rho(x_i,t)=\rho(r,t)$ and $\vec v(x_i,t)=\frac{\vec
x}{r}v(r,t)$. Substituting this parametrization into equations
(\ref{8}) and (\ref{17}) we obtain:
\begin{eqnarray}\label{25}
&\dot{v}(r,t)+v(r,t)\partial_r v(r,t)=\gamma \partial _r \int d^3x'\frac{\rho(r',t)}{|\vec x-\vec
x'|}\\ \nonumber
&r^2\dot\rho(r,t)+\partial_r(r^2\rho(r,t)v(r,t))=0
\end{eqnarray}
As is well known for spherically symmetric mass distribution the
force in the r.h.s. of the first equation (25) is defined by the
total mass inside the sphere of radius $r$:
\begin{equation}\label{26}
\gamma \partial _r \int d^3x'\frac{\rho(r',t)}{|\vec x-\vec x'|}
=-4\pi\gamma\frac{m(r,t)}{r^2},
\end{equation}
where we introduced notation
\begin{equation}
m(r,t)=\int_0^r dr' r'^2\rho(r',t)
\end{equation}
With this notation we can present the equations (24) in the following form:
\begin{eqnarray}\label{27}
&\dot{v}(r,t)+v(r,t)\partial_r v(r,t)=-4\pi\gamma\frac{m(r,t)}{r^2}\\ \nonumber
&\dot {m}(r,t)+v(r,t)\partial_r m(r,t)=0.
\end{eqnarray}
Let us consider the function $f(r,t)$ which is defined by the
following equation:
\begin{equation}\label{28}
\dot {f}(r,t)+v(r,t)\partial_r f(r,t)=0, \qquad f(r,0)=r.
\end{equation}
The function $m(r,t)$, because it satisfies the same equation could
be written in the following form:
\begin{equation}\label{29}
m(r,t)=m(f(r,t)),
\end{equation}
where $m(r)$ is the initial data for function $m(r,t)$:
$m(r)=m(r,0)$. In such a way we can rewrite the first of equations
(27):
\begin{equation}\label{30}
\dot{v}(r,t)+v(r,t)\partial_r
v(r,t)=-4\pi\gamma\frac{m(f(r,t))}{r^2}.
\end{equation}
Multiplying  both sides of this equation by $v(r,t)$ and taking into account equation (28)
we arrive at the following statement:
\begin{equation}\label{31}
(\partial_t+v(r,t)\partial_r)[\frac{v(r,t)^2}{2}-4\pi\gamma\frac{m(f(r,t))}{r}]=0.
\end{equation}
From this statement follows that the expression in the brackets is a
function of $f(r,t)$ only. So, we have
\begin{equation}\label{32}
\frac{v(r,t)^2}{2}-4\pi\gamma\frac{m(f(r,t))}{r}=A(f(r,t)).
\end{equation}
Putting in (32) $t=0$ we can express the unknown function A(r)
via initial data for $v(r,t)$ and $m(r,t)$:
\begin{equation}\label{33}
\frac{v(r,0)^2}{2}-4\pi\gamma\frac{m(r)}{r}=A(r).
\end{equation}
Therefore we obtain
\begin{equation}\label{34}
\frac{v(r,t)^2}{2}=\frac{v(f(r,t))^2}{2}+4\pi\gamma
m(f(r,t))[\frac{1}{r}-\frac{1}{f(r,t)}],
\end{equation}
where we put $v(r,0)=v(r)$. Apparently from (34) we can find the
velocity:
\begin{equation}\label{35}
v(r,t)=[2(\frac{v(f(r,t))^2}{2}+4\pi\gamma
m(f(r,t))[\frac{1}{r}-\frac{1}{f(r,t)}])]^{1/2}.
\end{equation}
This expression for velocity has to be substituted into equation (28)
for function $f(r,t)$ which we need to solve. In order to do it let
us introduce the function $s(r,t)$:
\begin{equation}\label{36}
r=f(r,t)-s(f(r,t),t).
\end{equation}
It is obvious that $s(r,0)=0$. Also from this definition follows
that
\begin{equation}\label{37}
v(r,t)=-\partial_t s(f,t).
\end{equation}
where we deliberately suppressed the arguments of the function $f(r,t)$
to make it clear that the differentiation in (37) is with respect to the
second argument of $s(f,t)$. (This relation could be obtained by applying
operator $\partial_t+v(r,t)\partial_r$ to both sides of (36)). In
such a way we have
\begin{equation}\label{38}
\partial_t s(f,t)=-[2(\frac{v(f)^2}{2}+4\pi\gamma
m(f)[\frac{1}{f-s(f,t)}-\frac{1}{f}])]^{1/2}.
\end{equation}
This relation gives us a possibility to find time dependence of the
function $s(f,t)$:
\begin{equation}\label{39}
-\int_0^{s(f,t)}\frac{ds}{[2(\frac{v(f)^2}{2}+4\pi\gamma
m(f)[\frac{1}{f-s}-\frac{1}{f}])]^{1/2}}=t
\end{equation}
Knowing the function $s(f,t)$ we can find function $f$ making use of
Lagrange formula for analytic branch of inverse function
\cite{hermite}. In the present context this formula takes the
following form (here we will suppress the argument $t$ of the function $s(f,t)$, considering it as a
parameter). So we have the following relation:
\begin{equation}\label{40}
r=f-s(f)
\end{equation}
The inverse function $f(r)$ is given by
\begin{equation}\label{41}
f=r+s(r)+\frac{1}{2}(s^2(r))'+\frac{1}{3!}(s^3(r))''+\frac{1}{4!}(s^4(r))^{(3)}+
\cdots+\frac{1}{n!}(s^n(r))^{(n-1)}\cdots.
\end{equation}
This formula gives the complete solution of our problem, though in
that form it is not at all clear what is the evolution even for the
simplest cases. As an example we shall consider the case where we
can at least qualitatively describe evolution. The first simplification
we shall do is to put $v(f)=0$. In this case the equation (39) takes
the following form:
\begin{equation}\label{42}
-[8\pi\gamma m(f)]^{-1/2}\int_0^{s(f,t)}\frac{ds}{
[\frac{1}{f-s}-\frac{1}{f}])]^{1/2}}=t.
\end{equation}
Now we shall change the integration variable (this parametrization is usual
for Kepler problem \cite{arnold})
\begin{equation}\label{43}
s=f(1-\frac{1+cos\alpha}{2}),
\end{equation}
after which we obtain
\begin{equation}\label{44}
-\Bigl[\frac{f^3}{8\pi\gamma m(f)}\Bigr]^{1/2}\frac{1}{2}[\alpha+sin\alpha]=t.
\end{equation}
Next simplification consists in choice of the function $m(r)$. If we
take uniform initial density in the whole space $\rho(r)=\rho_0$,
then $m(r)=\rho_0\frac{r^3}{3}$ and equation(44) becomes
\begin{equation}\label{45}
-\Bigl[\frac{3}{8\pi\gamma\rho_0}\Bigr]^{1/2}\frac{1}{2}[\alpha+sin\alpha]=t.
\end{equation}
Let us summarize the result. The equation (36) gives us the relation
\begin{equation}\label{46}
r=f\frac{1+cos\alpha}{2}.
\end{equation}
Time dependence of $\alpha$ is given by equation (45), which we can
rewrite in the following form
\begin{equation}\label{47}
\alpha+sin\alpha=-2\pi\frac{t}{T},
\end{equation}
where we introduced notation
$T=\pi\Bigl[\frac{3}{8\pi\gamma\rho_0}\Bigr]^{1/2}$. Note that in
this case we don't need to use Lagrange formula in order to find the
function $f(r,t)$. From equation (46) we have
\begin{equation}\label{48}
f(r,t)=\frac{2r}{1+cos\alpha(t)},
\end{equation}
from where we obtain the evolution of the functions $m(r,t)$ and
$v(r,t)$:
\begin{eqnarray}\label{49}
& m(r,t)=m(f(r,t))=\frac{1}{3}\rho_0\Bigl[\frac{2r}{1+cos\alpha(t)}\Bigr]^3\nonumber\\
& v(r,t)=-\frac{\partial_t f}{\partial_r f}=
[\frac{8\pi\gamma\rho_0}{3}]^{1/2}\frac{2rsin\alpha(t)}{(1+cos\alpha(t))},
\end{eqnarray}
Also from $m(r,t)$ we obtain the evolution of density:
\begin{equation}\label{50}
\rho(r,t)=\frac{1}{r^2}\partial_r
m(r,t)=\rho_0\Bigl[\frac{2}{1+cos\alpha(t)}\Bigr]^3.
\end{equation}
In all these equations time dependence of $\alpha(t)$ is defined by
equation (47). Note that according to it $\alpha(t)$ is zero at
$t=0$ and moves in negative direction. We see the the density is
unform for all $t$, oscillating between  $\rho_0$ and infinity.

\section{Evolution of flat configuration gravitating gas.}

Let us consider the case where $\vec v=(v(x,t),0,0)$ and $\rho=\rho(x,t)$. Euler equations takes
the following form:
\begin{eqnarray}\label{51}
&\displaystyle\dot{v}(x,t)+v(x,t) \partial_x v(x,t)=-2\pi\gamma \partial_x \int dy|x-x'|\rho(x',t)\nonumber\\
&\dot{\rho}(x,t)+\partial_x(\rho(x,t)v(x,t))=0
\end{eqnarray}
This form could be obtained from the 3D equations and the kernel $-2\pi|x-x'|$ is the result of
integrating  $\frac{1}{|\vec x-\vec x'|}$ over variables which don't enter into $\rho$. Needless
to say that this one-dimensional gravitating gas could be interpreted as flat layer in three
dimension. Differentiating potential in the r.h.s of Euler equation we can rewrite it in the
following form:
\begin{eqnarray}\label{52}
&\displaystyle\dot{v}(x,t)+v(x,t) \partial_x v(x,t)=-2\pi\gamma \int dy\epsilon(x-x')\rho(x',t)\nonumber\\
&=-2\pi\gamma[\int_{-\infty}^x dx'\rho(x',t)-\int_x^{\infty}dx'\rho(x',t)],
\end{eqnarray}
where $\epsilon(x)=sign(x)$.
It will be convenient to introduce instead of density another variable $g(x,t)$
\begin{equation}\label{53}
g(x,t)=\frac{1}{2}\int dy\epsilon(x-x')\rho(x',t)=
\frac{1}{2}(\int_{-\infty}^x dx'\rho(x',t)-\int_x^{\infty}dx'\rho(x',t)).
\end{equation}
Apparently this new variable plays the same role as the function
$m(r,t)$ in the previous section. With this new variable the the
system (\ref{51}) will take the following form:
\begin{eqnarray}\label{54}
&\dot{v}(x,t)+v(x,t) \partial_x v(x,t)=-4\pi\gamma g(x,t)\nonumber\\
&\dot{g}(x,t)+v(x,t) \partial_x g(x,t)=0.
\end{eqnarray}
Now let us introduce the first integral of the second equation $f(x,t)$,
satisfying initial condition $f(x,0)=x$. Apparently, the function $g(x,t)$ could be written as
\begin{equation}\label{55}
g(x,t)=g(f(x,t),0),
\end{equation}
so, evolution of $g(x,t)$ is just reparametrization of its initial
data $g(x,0)=g(x)$. Now it is easy to find general solution of the
first equation (\ref{54}):
\begin{equation}\label{56}
v(x,t)=v(f(x,t))-t4\pi\gamma g(f(x,t)),
\end{equation}
where $v(x)$ is the initial data for velocity. The first term in
(\ref{56}) is the solution of homogenous equation which we fixed
through the initial data for $v(r,t)$, while the second is the
solution of inhomogeneous one. Now we have arrived at the point
similar to that in the previous section. We need to find the
function $f(x,t)$. We shall do it as before. Let us introduce the
function $s(f,t)$ through the equation
\begin{equation}\label{57}
x=f(x,t)-s(f(x,t),t),
\end{equation}
such that
\begin{equation}\label{58}
\partial_t s(f,t)=-v(x,t),
\end{equation}
where the derivative is taken with respect to the second argument.
The case, we are considering now is simpler when the previous one
and we can find $s(f,t)$ explicitly:
\begin{equation}\label{59}
s(f,t)=-t v(f)+4\pi\gamma\frac{t^2}{2}g(f).
\end{equation}
Now we again can use Lagrange formula to find $f(x,t)$
\begin{equation}\label{60}
f=x+s(x)+\frac{1}{2}(s^2(x))'+\frac{1}{3!}(s^3(x))''+\frac{1}{4!}(s^4(x))^{(3)}+
\cdots+\frac{1}{n!}(s^n(x))^{(n-1)}\cdots.
\end{equation}
Remember that here we suppressed the second argument of the function
$s(f,t)$, considering $t$ as a parameter.

So, formally the Euler equations are solved if we can evaluate for
given initial data $\rho_0 (x),v(x)$ the function $f(x,t)$ presented
by infinite series (\ref{60}). Again, for some simple  initial data
it is possible to solve equation (\ref{59}) without Lagrange
formula. For example let us take $v(x)=0$ and density $\rho_0(x)=b$
for $|x|\leq a$ and $\rho_0(x)=0$ outside this interval. In this
case the function $s(f)$ in (\ref{59}) is linear and we can find
explicitly evolution of gravitating gas.
\begin{eqnarray}\label{61}
g(f(x,t))=\cases{-ab,& $x< -a(1-2\pi\gamma t^2 b)$,\cr
\frac{bx}{(1-2\pi\gamma t^2 b)},& $-a(1-2\pi\gamma t^2 b)\leq x \leq a(1-2\pi\gamma t^2 b)$,\cr
ab,& $x > a(1-2\pi\gamma t^2 b)$.\cr}
\end{eqnarray}
Differentiating $g(f(x,t))$ by $x$ we obtain from (\ref{34}) evolution of density
\begin{eqnarray}\label{62}
\rho(x,t)=\cases{0,& $x< -a(1-2\pi\gamma t^2 b)$,\cr
\frac{b}{(1-2\pi\gamma t^2 b)},& $-a(1-2\pi\gamma t^2 b)\leq x \leq a(1-2\pi\gamma t^2 b)$,\cr
0,& $x > a(1-2\pi\gamma t^2 b)$\cr}
\end{eqnarray}
and from (\ref{58}) we can get also evolution of velocity. This
solution of Euler equations which we obtained describes collapse of
gravitating gas during finite time $T=\sqrt{\frac{1}{2\pi\gamma b}}$
to zero size. Here arises the question what happens after this time.
The analysis of the Lagrange equation shows that the motion
continues in backward evolution. Gas emerges from one point and
returns to the initial configuration. The impossibility of analytic
description of the whole evolution has the same origin as in the
problem of bouncing ball.

\section{Acknowledgments}

I would like to thank my friends professors A. Likhoded and A.
Rasumov for their interest and support. The work on this paper was
supported in part by the Russian Science Foundation Grant
10-01-00300.

\end{document}